\documentstyle[12pt]{article}
 \newcommand{\bea}{\begin{equation}}
 \newcommand{\eea}{\end{equation}}
 \newcommand{\ber}{\begin{eqnarray}}
 \newcommand{\eer}{\end{eqnarray}}
\begin{document}
\title{SCALING FUNCTION FOR THE CRITICAL DIFFUSION COEFFICIENT OF A CRITICAL FLUID IN A FINITE GEOMETRY}
\author{Palash Das and Jayanta K.Bhattacharjee\\
Department of Theoretical Physics \\
Indian Association for the Cultivation of Science \\
Jadavpur, Calcutta 700 032, India}
\date{}
\maketitle
\begin{abstract}.
The longwavelength diffusion coefficient of a critical fluid confined between
two parallel plates seperated by a distance L is strongly affected by the
finite size. Finite size scaling leads us to expect that the vanishing of the
diffusion coefficient as $\xi^{-1}$, for $\xi\ll L$, $\xi$ being the
correlation length, would crossover to $L^{-1}$ for $\xi\gg L$. We show that
this is not strictly true. There is a logarthmic scaling violation. We
construct a Kawasasi like scaling function that connects the thermodynamic
regime to the extreme critical ($\xi\gg L$) regime. \vspace{1cm}

PACS number(s):64.60Ht
\end{abstract}
\newpage
One of the earliest and certainly one of the most frequently used scaling
function in the whole of critical dynamics is the scaling function for the
thermal conductivity / concentration diffusivity in a single component / binary
fluid near the gas-liquid critical point / critical mixing point. This is the
well known Kawasaki function which describes the passage of the diffusion
coefficient (we will use the binary liquid language) from a long-wavelength,
finite correlation length  regime to a finite-wavelength, infinite correlation
length regime. When the wavelength is very large (k, the wave number
$\rightarrow 0$), the diffusion coefficient D diverges as
$\xi^{4-d}=\kappa^{d-4}$, where $\xi$ is the correlation length; $\kappa$, the
inverse correlation length and d, the dimensionality of space. At the critical
point ($\kappa=0$), this behaviour changes to $k^{d-4}$. In the physical (d=3)
situation, we have the passage described by the Kawasaki function $K(x)$ [1]:
\bea D(k,\kappa)=\frac{k_{B}T}{6\pi\eta_{0}\xi}K(k\xi) \eea with \bea
K(x)=\frac{3}{4x^{2}}[1+x^{2}+(x^{3}-\frac{1}{x})tan^{-1}x] \eea If
$k\rightarrow0$, $D(\kappa)=\frac{k_{B}T}{6\pi\eta_{0}\xi}$ and if
$\kappa\rightarrow0$, $D(k)=\frac{k_{B}T}{6\pi\eta_{0}}\frac{3\pi}{8}k$. In the
above $\eta_{0}$ is the background critical viscosity. The small divergence of
the critical viscosity is ignored in this paper. A practical approximation to
the Kawasaki function is to write it as

\bea D(k,\kappa)=\frac{k_{B}T}{6\pi\eta_{0}}\L \eea with $\L$ having the
property that \ber
\L&\rightarrow&\frac{1}{\kappa}\hspace{.1cm} as \hspace{.1cm} k\rightarrow0\nonumber\\
&\rightarrow&\frac{3\pi}{8}\frac{1}{k}=\frac{1.18}{k}\simeq\frac{1}{k}\hspace{.1cm} as \hspace{.1cm}\kappa\rightarrow 0\nonumber\\
& & {} \eer The first correction being known to be of the form
$\frac{k^{2}}{\kappa^{2}}$ about the $k=0$ limit, we can use the approximation
$\L(k,\kappa)=(k^{2}+\kappa^{2})^{-1}$ and the diffusion coefficient is \bea
D(k,\kappa)=\frac{k_{B}T}{6\pi\eta_{0}}(k^{2}+\kappa^{2})^{\frac{1}{2}} \eea
This is exact for $k\rightarrow 0$ and over the range of practical values of
the $\frac{k}{\kappa}$ ratio, is accurate to about $5\%$. This makes it a very
useful approximation.
\par
In this paper, we want to look at the diffusion coefficient in a finite
geometry - the fluid contained between two slabs separated by a distance L.
Recently Koch and Dohm have explored finite size effects on the diffusion
coefficient in three dimensional Ising-like systems [17]. Unlike their
situation, we will show that our system exhibits a small scaling violation.
This is a physical model and hence this violation should be experimentally
accessible. The hydrodynamic limit is taken to hold i.e. we are in the long
wavelength limit (i.e. k=0). For $L\gg\xi$, the diffusion coefficient will be
governed by $D(\kappa)=\frac{k_{B}T}{6\pi\eta_{0}}\kappa$. In the finite
geometry, the leading correction to this result was first obtained by Calvo and
Ferrell. The result was a bit of a surprise in the presence of a logarithmic
factor \bea \L(\kappa,L)=\kappa^{-1}[1-\frac{ln\kappa L}{2\kappa
L}],\hspace{.5cm} \kappa L \gg 1 \eea This was later obtained using a different
technique by one of the present authors [16]. The appearance of the logarithmic
term prompted us to explore the limit $\xi\gg L$. In this limit, we find (as
explained below)
\begin{displaymath}
\L(\kappa=0,L)=\frac{L}{12}ln\frac{1}{\kappa L}
\end{displaymath}
To arrive at the diffusion coefficient, one needs
\begin{displaymath}
D(\kappa,L)=\frac{k_{B}T}{6\pi\eta_{0}}\frac{\L(\kappa,L)}{\chi(\kappa,L)}
\end{displaymath}
where $\chi(\kappa,L)$ is the susceptibility in the finite geometry. The
limiting form of $\chi(\kappa,L)$ are $\chi(\kappa, L\rightarrow
\infty)=\kappa^{-2}$ and $\chi(\kappa\rightarrow 0,L)=\frac{L^{2}}{12}$, so
that\ber
D(k,\kappa)&\rightarrow&\frac{k_{B}T}{6\pi\eta_{0}}\kappa (1-\frac{1}{2}\frac{ln\kappa L}{\kappa L}),\hspace{.1cm} for \hspace{.1cm}\kappa^{-1} < L\nonumber\\
&\rightarrow& \frac{k_{B}T}{6\pi\eta_{0}}\frac{1}{L}ln\frac{1}{\kappa L},\hspace{.1cm} for \hspace{.1cm}\kappa^{-1}\gg L \nonumber\\
& & {}\eer We combine the two forms to propose the scaling function \ber
D(k,\kappa)&=&\frac{k_{B}T}{6\pi\eta_{0}\xi}(1+\frac{1}{\kappa^{2}L^{2}})\frac{1+\frac{1}{8}[ln(1+\frac{1}{\kappa^{2}L^{2}})]^{2}}{1+\frac{1}{2\kappa L}ln(\kappa L+\frac{1}{\kappa L})}\nonumber\\
&=& \frac{k_{B}T}{6\pi\eta_{0}\xi}F(\kappa L)\nonumber\\
& & {}\eer which is the ($\kappa,L$) analogue of the Kawasaki scaling function,
given in Eq.(2). \par In the thermodynamic limit ($\kappa L \rightarrow
\infty$), $F(\kappa L)=1$ and rises to $\frac{1}{\kappa L}ln\frac{1}{\kappa L}$
as $\kappa L \rightarrow 0$. The change in $\kappa L$ is brought about by
varying $\xi$ at a fixed L. At $\kappa L =1$, $F(\kappa L)=1.57$, significantly
different from its value in the thermodynamic limit. This should make the
effect observable. To see the existence of the logarithmic terms, one would
need a fairly high degree of accuracy. We now proceed to outline the technical
details.
\par
The free energy functional that governs the static fluctuation of the order
parameter $\psi$ can be taken to be quadratic and is given by \ber F&=&\int
d^{D-1}r\int_{0}^{L}dz[\frac{\kappa^{2}}{2}\sum_{i=1}^{n}\psi_{i}^{2}+\frac{1}{2}\sum_{i=0}^{n}(\vec
\nabla \psi_{i})^{2}]+\nonumber\\
& & \frac{c}{\lambda}\int d^{D-1}r [\psi^{2}(\vec r,z=0)+\psi^{2}(\vec r, z=L)]\nonumber\\
& & {} \eer Since the anomalous dimension index $\eta$ plays an insignificant
role in the study of dynamics, we can work with this quadratic expression for
$F$. Here the geometry is restricted in the z-direction between $z=0$ and
$z=L$. The second term on the right hand side of Eq.(9) is a surface
contribution, 'c' is a constant and $\lambda$, an extrapolation length. For
$\lambda \rightarrow 0$, $\psi^{2}$ must vanish at $z=0$ and $z=L$ in order to
satisfy the Dirichlet boundary conditions. We set $\lambda=0$.
\par
We introduce the Fourier transform of $\psi(\vec r,z)$ through the relation
\bea \psi(\vec r,z)=\frac{1}{(2\pi)^{2}}\int \psi(\vec k,z)e^{i\vec k \cdot
\vec r} d^{2}k \eea where $\vec k$ is the wave-vector in the 2-dimensional
space. The two-point correlation function $G(\vec k, z_{1},z_{2})$ is the
solution of the differential equation \bea
(\frac{d^{2}}{dz^{2}}-k^{2}-\kappa^{2})G(k,z_{1},z_{2})=-\delta(z_{1}-z_{2})\eea
with $G(\vec k,z_{1},z_{2})$ vanishing at $z=0$ and $z=L$ and is given by \bea
G(\vec k,z_{1},z_{2})=\frac{sinhaz_{<}sinha(L-z_{>})}{2asinhaL}
 \eea
with $a^{2}=k^{2}+\kappa^{2}$. For $L\rightarrow 0$ and $z_{1}>z_{2}$, Eq.(12)
approximates to \bea G\sim \frac{z_{<}(L-z_{>})}{2L} \eea At this point it is
easy to check that $\chi(\kappa,L)=\frac{1}{L}\int G(\vec
k=0,z_{1},z_{2})dz_{1}dz_{2}$ gives the results mentioned above Eq.(7).
\par
Let us now focus our attention on dynamics. For this we need to introduce the
equation of motion for $\psi$-field. We take this to be a Langevin equation
where the potential corresponds to the free energy functional of Eq.(9). For
non-conserved $\psi$-field, this reads as \bea \frac{\partial \psi}{\partial
t}=-\Gamma (k^{2}+\kappa^{2}-\frac{\partial^{2}}{\partial z^{2}})\psi(\vec
k,z,t)+N(\vec k,z,t)\eea where $N(\vec k,z,t)$ is the noise, characterized by
the correlation \bea <N(\vec k_{1},z_{1},t_{1})N(\vec
k_{2},z_{2},t_{2})>=2\Gamma \delta(\vec k_{1}+\vec
k_{2})\delta(z_{1}-z_{2})\delta(t_{1}-t_{2})\eea If the order parameter field
$\psi$ is conserved, there will be an additional
factor\\$(k^{2}-\frac{\partial^{2}}{\partial z^{2}})$ multiplying $\Gamma$ in
Eq.(14) [2]. The dynamic correlation function reads as \ber C(\vec
k,z_{1},z_{2})&=&<\psi(\vec k,z_{1},\omega)\psi(-\vec
k,z_{2},-\omega)>\nonumber\\
&=& \frac{1}{\Gamma^{2}}\int
dz^{\prime}dz^{\prime\prime}R_{+}(z_{1},z^{\prime})R_{-}(z_{2},z^{\prime\prime})<N(z^{\prime})N(z^{\prime\prime})>\nonumber\\
&=&\frac{2}{\Gamma}\int dz^{\prime}
R_{+}(z_{1},z^{\prime})R_{-}(z_{2},z^{\prime})\nonumber\\
& & {} \eer where \bea
R_{\pm}=\frac{sinha_{\pm}z_{<}sinha_{\pm}(L-z_{>})}{2a_{\pm}sinha_{\pm}L} \eea
with \bea a^{2}_{\pm}=\pm \frac{i\omega}{\Gamma}+k^{2}+\kappa^{2} \eea
\par
 Now, we note that the current $\vec j(\vec r,z,t)$ associated
with a transport process is in general a bilinear combination of the order
parameter field or a combination of the order parameter field with a secondary
field. Then Kubo's formula yields the Onsager coefficient corresponding to the
current $\vec j(\vec r,z,t)$ as \bea \lambda \propto \frac{1}{V_{2}L T}\int
<\vec j(\vec r_{1},z_{1},t_{1}) \cdot \vec j(\vec
r_{2},z_{2},t_{2})>d^{2}r_{1}d^{2}r_{2}dz_{1}dz_{2}dt_{1}dt_{2} \eea where
$V_{2}$ is the volume in the two dimensional space.
\par
For the binary liquid system, the order parameter field $\psi$ is the density
difference between the liquid and gaseous phase and the relevant current is
$\vec j = \psi \vec v$, where $\vec v$ is the velocity field. For the
liquid-gas system, this current is proportional to the entropy current and so
Kubo's formula yields the thermal diffusivity. For a binary mixture the current
is the mass current and Kubo's formula yields the mass diffusivity. Using $\vec
j=\psi \vec v$ in Eq.(19), we have \ber
 \lambda &\propto& \frac{1}{V_{2}L T}\int <\psi(\vec r_{1},z_{1},t_{1})\vec v(\vec r_{1},z_{1},t_{1}) \cdot \psi(\vec r_{2},z_{2},t_{2})\vec v(\vec r_{2},z_{2},t_{2})>\nonumber\\
& & \hspace{3cm}\times d^{2}r_{1}d^{2}r_{2}dz_{1}dz_{2}dt_{1}dt_{2}\nonumber\\
& & {} \eer Then the decoupled mode approximation enables us to write the
correlation function in Eq.(20) as a product of two
correlation functions [3] viz. \\
$<\psi(\vec r_{1},z_{1},t_{1})\psi(\vec r_{2},z_{2},t_{2})>$ and $<\vec v(\vec
r_{1},z_{1},t_{}1) \cdot \vec v(\vec r_{2},z_{2},t_{2})>$ and so finally we
have \bea \lambda \propto \frac{1}{L}\int C_{\psi\psi}(\vec
k,z_{1},z_{2},\omega)C_{vv}(-\vec k,z_{1},z_{2},\omega)d^{2}k d\omega
dz_{1}dz_{2}\eea where $C_{\psi\psi}$ and $C_{vv}$ respectively stand for the
order parameter and velocity correlation functions.
\par

We now note that the time scales associated with the velocity field and the
density field are very different. The velocity field relaxes much faster and in
the time scale $(t_{1}-t_{2})$, the density field changes hardly. As a result
the order parameter correlation function $C_{\psi\psi}(\vec r_{1}-\vec
r_{2},z_{1},z_{2},t_{1}-t_{2})$ can be taken to remain at its static value and
hence one needs the zero frequency limit of the velocity correlation function
$C_{vv}$. Therefore Eq.(21) reduces to \bea \lambda \propto \frac{1}{L}\int
d^{2}k dz_{1}dz_{2}C{\psi\psi}^{static}(\vec k,z_{1},z_{2})C_{vv}(-\vec
k,z_{1},z_{2},\omega=0) \eea
\par
The static correlation function for the order parameter $\psi$ is given by
Eq.(2). Let us now proceed to evaluate the zero frequency limit of the velocity
correlation function $c_{vv}(-\vec k,z_{1},z_{2},\omega)$. The relaxation
dynamics of the $\vec v$-field is governed by \bea \frac{\partial \vec
v}{\partial t}=-\Gamma_{v}(k^{2}-\frac{\partial^{2}}{\partial z^{2}})\vec
v(\vec k,z,t)+\vec N_{v} \eea The velocity field is solenoidal and so $\vec
\nabla \cdot \vec v =0$. The above constraint tells us that we should work with
a field $\vec A$, such that $\vec v=\vec \nabla \times \vec A $. The
correlation function $C_{AA}$ for the $\vec A$-field is given by Eq.(12) with\\
$\kappa=0\Rightarrow T=T_{c}$. The velocity correlation function follows from
\bea C_{vv}=(k^{2}+\frac{\partial^{2}}{\partial z_{1}\partial z_{2}})C_{AA}
\eea  In the limit $L \rightarrow 0$, it is found to be \bea
C_{vv}=\frac{1}{2\Gamma}[\frac{1}{2k^{2}L^{2}}(4z_{2}-z_{1})] \eea
\par
Using Eqs.(13) and (26), the Onsager coefficient, in the three spatial
dimensions, is \ber \lambda &\propto& \frac{1}{L}\int d^{2}k
dz_{1}dz_{2}C_{\psi\psi}^{static}(\vec k,z_{1},z_{2})C_{vv}(-\vec k,z_{1},z_{2},\omega=0)\nonumber\\
&=& \frac{1}{L}\int d^{2}k \int_{0}^{L}[\int_{0}^{z_{1}}+\int_{z_{1}}^{L}]dz_{2}\nonumber\\
& & \times \frac{1}{4\Gamma
k^{2}L^{2}}(4z_{2}-z_{1})\frac{z_{2}(L-z_{1})}{2L},\hspace{1cm}
(z_{1}>z_{2})\nonumber\\
&=&\frac{1}{4\Gamma L^{4}}\int
\frac{d^{2}k}{k^{2}}\int_{0}^{L}(L-z_{1})dz_{1}\int_{0}^{z_{1}}
z_{2}(4z_{2}-z_{1})dz_{2},\nonumber\\
& & since\hspace{.5cm}\int_{0}^{L}dz_{1}[\int_{0}^{z_{1}}+\int_{z_{1}^{L}}]dz_{2}=2\int_{0}^{L}dz_{1}\int_{0}^{z_{1}}dz_{2}\nonumber\\
&=& \frac{1}{4\Gamma L^{4}}\frac{5}{6}\int
\frac{d^{2}k}{k^{2}}\int_{0}^{L}z_{1}^{3}(L-z_{1})dz_{1}\nonumber\\
&=& \frac{L}{96\Gamma}\int \frac{d^{2}k}{k^{2}}\nonumber\\
& & {} \eer Therefore, for $L \rightarrow 0$ \bea \lambda \propto
L\int_{\kappa}^{L^{-1}}\frac{dk}{k}\sim L ln\frac{1}{\kappa L} \eea This
establishes the point we wanted to make in Eq.(7). Normalizing to the same
prefactor for the large $\kappa$ and small $\kappa$ limits leads to the
coefficients in Eq.(7).

%\noindent$ \bf {Figure\:\:caption} $ \\\\


\newpage
\begin{thebibliography}{99}
\bibitem[1]{kawa}
K.Kawasaki Ann.Phys.(N.Y.). 61. 1. (1970).
\bibitem[2]{cross}
P.C.Hohenberg and B.I.Halperin, Rev.Mod.Phys. {\bf 49}, 435, (1977).
\bibitem[3]{win}
R.A.Ferrell, Phys.Rev.Lett. {\bf 24}, 1169, (1970).
\bibitem[4]{kap}
J.P.Chen and F.M.Gasparini, Phys.Rev.Lett. {\bf 40}, 331, (1978).
\bibitem[5]{bel}
J.A.Nissen and T.C.P.Chui and J.A.Lipa, J.Low Temp.Phys. {\bf 92}, 353, (1993).
\bibitem[6]{nishi}
I.Rhee, F.M.Gasparini and D.J.Bishop, Phys.Rev.Lett. {\bf 63}, 410, (1989).
\bibitem[7]{dur}
G.Ahlers and R.V.Duncan, Phys.Rev.Lett. {\bf 61}, 846, (1988).
\bibitem[8]{vas}
A.M.Kahn and G.Ahlers, Phys.Rev.Lett. {\bf 73}, (1995).
\bibitem[9]{kess}
W.Huhn and V.Dohm, Phys.Rev.Lett. {\bf 61}, 1368, (1988).
\bibitem[10]{winf}
V.Dohm, Physics Scripta {\bf T49}, 49, (1993).
\bibitem[11]{tu}
P.Suiter and V.Dohm, Physica B, (1993).
\bibitem[12]{mar}
R.Schmolke, A.Wacker, V.Dohm and D.Frank, Physica {\bf B165} and {\bf B166},
575, (1990).
\bibitem[13]{ott}
M.Krech and S.Dietrich, Phys.Rev.Lett. {\bf 66}, 345, (1992).
\bibitem[14]{hagan}
H.L.Swinney and D.L.Henry, Phys.Rev.A. {\bf 8}, 2586, (1973).
\bibitem[15] {abc}
K.Fossheim and J.O.Fossum, Invited lecture at NATO ASI on "Multicritical
Phenomena", Geilo, Norway, 1983.
\bibitem[16] {jay}
J.K.Bhattacharjee, Phys.Rev.Lett. {\bf 77}, 1524, (1996).
\bibitem[17] {bal}
W.Koch and V.Dohm, Phys.Rev.E. {\bf 58}, R1179, (1998) and also see X.S.Chen,
V.Dohm and N.Schultka, Phys.Rev.Lett. {\bf 77}, 3641, (1996) and W.Koch, V.Dohm
and D.Stauffer, Phys.Rev.Lett. {\bf 77}, 1789, (1996).






















\end{thebibliography}
\end{document}